\documentclass[12pt]{article}
\usepackage{graphics}
\usepackage{epsfig}
\usepackage{hyperref}
\hypersetup{
  pdftitle= ,
  pdfsubject= ,
  pdfauthor=Dr. Mulugeta ,
  pdfmenubar=true,
  pdftoolbar=true,
  pdfpagemode={None},
  colorlinks=false,
  bookmarks=true,
  pdfkeywords= Water,
  }
\usepackage{graphics}
\usepackage{graphicx}
\title{\bf Adhesion-induced lateral phase separation of multi-component membranes:
the effect of repellers and confinement}
\author{ Mesfin Asfaw and Hsuan-Yi Chen \\
        Department of Physics and Institute of Biophysics\\
        National Central University,  Jhongli 32001, Taiwan }
\date{Received: date / Revised version: date}
\begin{document}
\maketitle

\begin{abstract}
We present a theoretical study for adhesion-induced lateral phase
separation for a membrane with short stickers, long stickers and
repellers confined between two hard walls. The effects of
confinement and repellers on lateral phase separation are
investigated. We find that the critical potential depth of the
stickers for lateral phase separation increases as the distance
between the hard walls decreases. This suggests confinement-induced
or force-induced mixing of stickers.  We also find that stiff
repellers tend to enhance, while soft repellers tend to suppress
adhesion-induced lateral phase separation.
\end{abstract}
\newpage
\section{Introduction}
Biological membranes are lipid bilayers with different types of
embedded or absorbed macromolecules. They serve a number of general
functions in our cells and tissues \cite{h1,h2}. Because of its
biological importance, the physics of membrane adhesion has received
considerable attention both theoretically and experimentally
\cite{h3,h4,h5,h6,h7,h8,h9,h11}. For instance, helper T cells
mediate immune responses by adhering to antigen-presenting cells
(APCs) which exhibit foreign peptide fragments on their
surface.~\cite{h12}. The APC membranes contain the ligands MHCp  and
ICAM-1 while   the   T cells contain  the receptors   TCR and LFA-1.
The experiments  \cite{h12} show  the formation  of domains into
shorter TCR/MHCp receptor-ligand complexes and the longer
LFA-1/ICAM-1 receptor-ligand complexes. The dynamics of
adhesion-induced  phase separation has been studied theoretically
\cite{h13,h14,h15}. For example, the Monte Carlo study by  Weikl and
Lipowsky \cite{h15} shows that the height difference between
different junctions causes a lateral phase separation, and the
formation of target-like immunological synapse is assisted by the
motion of cytoskeleton.

The equilibrium studies of adhesion-induced phase separation of
multi-component membranes are also important for a complete
understanding of the physics of membrane adhesion. For instance, in
recent articles \cite{h16, h17}, the general case of two membranes
binding to each other with two types of stickers is considered and
the equilibrium phase behavior of such a system is studied at the
mean field and Gaussian level by including the effects of sticker
flexibility difference, sticker height difference and thermally
activated membrane height fluctuations. More recently, Mesfin $et$
$al.$ \cite{h18} presented a theoretical study that characterized
the phase diagram and the scaling laws for the critical potential
depth of unbinding and lateral phase separation. These studies show
that membranes are unbound for small potential depths and bound for
large potential depths. In the bound state, the length mismatch
leads to a membrane-mediated repulsion between stickers of different
lengths and this leads to lateral phase separation depending the
concentrations and strengths of the receptor-ligand bonds.
Furthermore, the flexibilities of the stickers play non-trivial
roles in the location of phase boundaries.

Most of these recent works deal with membranes with one or two types
of stickers. However, biological membranes usually contain
glycoproteins which are repulsive to another membrane or tissue,
i.e., they act as repellers.  This important fact motivates us to
study adhesion-induced lateral phase separation of membranes with
short stickers, long stickers and repellers. Another important but
unexplored issue on adhesion-induced lateral phase separation in
biomembranes is the effect of external pressure or confinement on
the phase diagram. For example, cell adhesions often occur in the
presence of external force field due to external flow, or the
external force may be a result of the occurrence of cell adhesions
in highly confined geometry during the development of multicellular
organisms. To study the effect of repellers and confinement on
adhesion-induced lateral phase separation, in this article we first
consider a membrane with short stickers and long stickers which are
in contact with another planer surface (substrate) in the absence of
repellers. The membrane and the substrate are confined between two
hard walls. We find that the critical binding energies of the
stickers for lateral phase separation increase as the distance
between the hard walls decreases due to the steric repulsion of the
membrane with the hard walls. Then the effect of repellers are
considered and we find that stiff repellers tend to enhance phase
separation, while soft repellers tend to suppress phase separation.
Our study has revealed the possibility to manipulating the lateral
distribution of stickers in future experiments.

This paper is organized as follows:  In section II we present the
model of membranes with short stickers, long stickers and repellers.
By tracing out sticker and repeller degrees of freedom, we get
membranes that interact  with an effective double-well potential.
The adhesion-induced lateral phase separation in the presence of
stickers and repellers is studied by mean field theory and Monte
Carlo simulations in section III. First we consider membranes with
short and long stickers. We then consider membranes with short, long
stickers and repellers. Section IV is the summary and conclusion.

\section{The model }
    We consider a tensionless non-homogenous multi-component membrane
with short and long receptor-ligand bonds that interacts with a
substrate as shown in Fig. 1. Let us denote the short and long
receptor-ligand bonds as short and long stickers, respectively. In
our model, the membrane is discretized into a two dimensional square
lattice with lattice constant $a$  \cite{h8,h18}. The lattice
constant $a$ is chosen to be $a=6{\rm nm}$, the smallest length
scale for membrane continuum elasticity theory to be valid. The
separation field $l \geq 0$ describes the vertical distance between
the membrane and the substrate. An additional field $n_{i}=0, \ 1,\
2$, or $3$ denotes the occupation state of the $i$th site. $n_{i}=0$
indicates the absence of stickers and repellers at lattice site $i$
while $n_{i}=1 (2)$ denote the presence of a type-1(2) sticker at
lattice site $i$; $n_{i}=3$ denotes the presence of a repeller at a
site $i$.
\begin{figure}[h]
\begin{center}
\epsfig{file=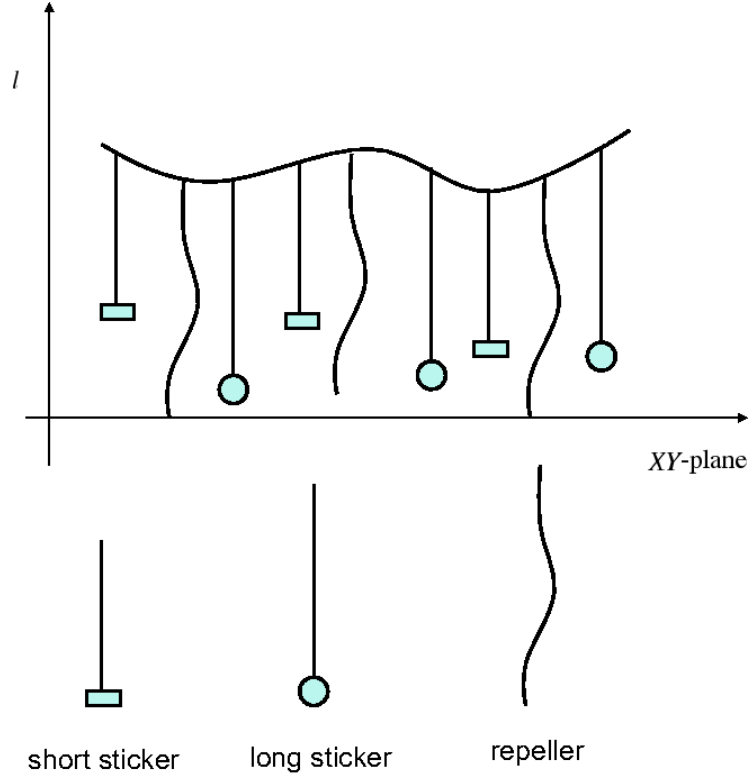,width=10cm} \caption{Schematic figure for
 a membrane with short stickers, long stickers and repellers close
 to a substrate. The local separation field is $l$.}
%\label{figfactor}
\end{center}
\end{figure}

The grand canonical Hamiltonian of the system under consideration is
given by
\begin{eqnarray}
         H[l,n]&=&H_{el}[l]+   \sum_{i} \delta_{1,n_i}(V_{1}(l_i)-\mu_{1})+ \nonumber \\&&
       \sum_{i} \delta_{2,n_i}(V_{2}(l_i)-\mu_{2})+ \sum_{i} \delta_{3,n_i}(V_{3}(l_i)-\mu_{3})
         \end{eqnarray}
here $H_{el}[l]= \sum_{i}{ \kappa \over 2a^2}(\Delta_{d}l_i)^2$
denotes the discretized bending energy of the membrane with bending
rigidity $\kappa $.  Typically, $\kappa = 10 - 20 {\rm k_{B}T}$. The
discretized Laplacian $\Delta_d $ is given by
$\Delta_{d}l_i=l_{i1}+l_{i2}+l_{i3}+l_{i4}-4l_{i}$ where $l_{i1}$ to
$l_{i4}$ are the four nearest-neighbor  membrane separation fields
of the membrane patch $i$. The second and third   terms on the right
hand side of Eq.~(1) are interaction potentials between the stickers
and the substrate.  $\mu_{1}$ and $\mu_{2}$ denote the chemical
potentials of  stickers 1 and 2, respectively. The parameters
$V_{3}({\it l}_{i})$  and $\mu_{3}$
represent potentials and the chemical potentials of the repellers, respectively.
We consider the following  sticker potentials: for $\alpha=1,2$,
\begin{eqnarray}
V_{\alpha} = \cases { U_{\alpha}, & if $
l_{\alpha}<l<l_{\alpha}+l_{we\alpha}$\cr 0, & otherwise \cr }
\end{eqnarray}
where $ U_{1}$, $U_{2}$ are both negative and $l_1 < l_2$.  That is,
type-1 stickers are shorter than type-2 stickers.  The repulsive
potential of the  repellers is  $ V_3 = U_{3} > 0$  for $0< l <
l_{3}$.

The equilibrium properties of the system can be obtained from the
grand partition function $Z$,
\begin{equation}
Z=\prod_{i} \int_{0}^{\infty }dl_{i} \sum_{n_{i}=0}^3
\exp\left[{-H[l,n]\over k_{B}T} \right].
\end{equation}
Absorbing the Boltzmann constant $k_{B}$ into the temperature $T$
and tracing out the sticker degrees of freedom one gets
\begin{eqnarray}
Z &=& \int_{0}^{\infty } \prod_{i} dl_{i}\exp
\left[-H_{el}(l)\right] \nonumber \\
&& \left[1+\exp \left[{-V_{1}(l_i)+\mu_{1}\over T}\right]+
\exp\left[{-V_{2}(l_i)+\mu_{2}\over T}\right]
+\exp \left[{-V_{3}(l_i)+\mu_{3}\over T}\right] \right] \nonumber \\
&=& \int_{0}^{\infty }dl_{i} \prod_{i} \exp
\left[{-H_{el}(l)+\sum_{i} V^{eff}(l_i) \over T}\right],
\end{eqnarray}
where the effective potential, $V^{eff}(l)$, is given by
\begin{equation}\label{eq:Veff}
V^{eff}=\cases{ %\infty,&for  $l  < 0$;\cr
             U_{ba},&for  $0< l  < l_{1}$;\cr
              U_{1}^{eff},&for  $l_{1}< l  < l_{1}+l_{we1}$;\cr
             U_{ba},&for  $l_{1}+l_{we1} < l< l_{2} ;$\cr
             U_{2}^{eff},&for  $l_{2} < l< l_{2}+l_{we2} ;$\cr
             U_{ba},&for  $l_{we2} < l< l_{3} ;$\cr 0,&otherwise,\cr
                    }
\end{equation}
where
\begin{equation}
U_{1}^{eff}=-{T} \ln{\left[1+\exp[{-U_{1}+\mu_{1}\over T}]+
            \exp[{\mu_{2}\over T}]+\exp[{\mu_{3}\over T}] \over 1+\exp[{\mu_{1}\over T}]+\exp[{\mu_{2}\over
            T}]+\exp[{\mu_{3}\over T}]\right]},
\end{equation}
\begin{equation}
U_{2}^{eff}=-{T} \ln{\left[1+\exp[{\mu_{1}\over T}]+
            \exp[{-U_2+\mu_{2}\over T}]+\exp[{\mu_{3}\over T}]\over 1+\exp[{\mu_{1}\over T}]+\exp[{\mu_{2}\over
            T}]+\exp[{\mu_{3}\over T}]\right]},
\end{equation}
and
\begin{equation}
U_{ba}^{eff}=-{T} \ln{\left[1+\exp[{\mu_{1}\over T}]+ \exp[{\mu_{2}\over T}]+\exp[{-U_{3}+\mu_{3}\over T}] %
\over %
1+\exp[{\mu_{1}\over T}]+\exp[{\mu_{2}\over T}]+ \exp[{\mu_{3}\over
T}]\right]},
           \end{equation}
as shown in Fig.~2.

In the following section, the phase behavior of membranes under the
effective potential given by Eq.~(\ref{eq:Veff}) will be studied by
mean field approximation and Monte Carlo simulations.
\begin{figure}[h]
\begin{center}
\epsfig{file=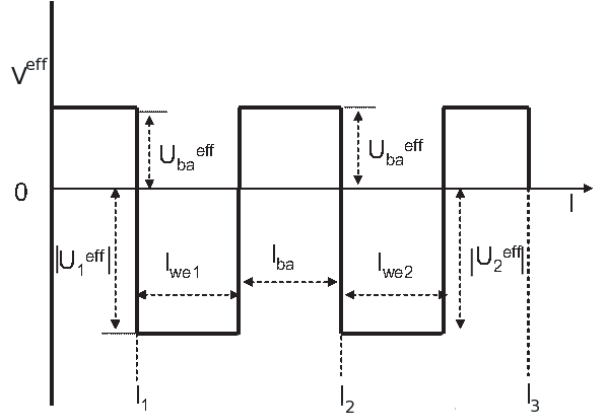,width=8cm} \caption{Schematic effective
potential, $V^{eff}$ versus  $l$. The potential has two square wells
of depths $|{\bar U}_{1}^{eff}|$  and $|{\bar U}_{2}^{eff}|$ and one
square barrier $U_{ba}^{eff}$. }
%\label{figfactor}
\end{center}
\end{figure}

\section{Mean field theory and Monte Carlo simulation}

It is convenient to introduce  the rescaled separation field
$z=(l/a)\sqrt{\kappa/T}$ and the rescaled effective potential ${\bar
V}^{eff}=V^{eff}/T$. In equilibrium state the rescaled separation
field $z$ fluctuates around its average value $z_{min}$. When the
fluctuation is not very  strong, mean field approximation can be
applied to  the discretized Laplacian such that  $ H_{l}[z]=
\sum_{i}(4[z_{min} -z_{i}])^2$.  In this approximation
 $z_i$ at different sites are decoupled.
Hence Eq. (4) becomes
\begin{eqnarray}
Z&=& \left[\int_{0}^{\infty }dz \exp{[-8(z_{min} -z)^2-{\bar
V}^{eff}(z)]}\right]^N,
\end{eqnarray}
and the mean field free energy of the membrane is given %after some algebra is given
by
\begin{eqnarray}
G&=&- NT \ln[\left[\int_{0}^{\infty }dz [\exp[-8(z_{min} -z)^2-{\bar
V}^{eff}(z)]]\right].
\end{eqnarray}
Minimizing the free energy (10) with  respect to $z_{min} $  leads
to the following self-consistence equation,
\begin{equation}
         z_{min} ={\int_{0}^{\infty }z\exp[-8(z_{min} -z)^2-{\bar
V}^{eff}(z)]dz \over \int_{0}^{\infty }\exp[-8(z_{min} -z)^2-{\bar
V}^{eff}(z)]dz}. \label{eq:zMFT}
\end{equation}
\begin{figure}[h]
\begin{center}
\epsfig{file=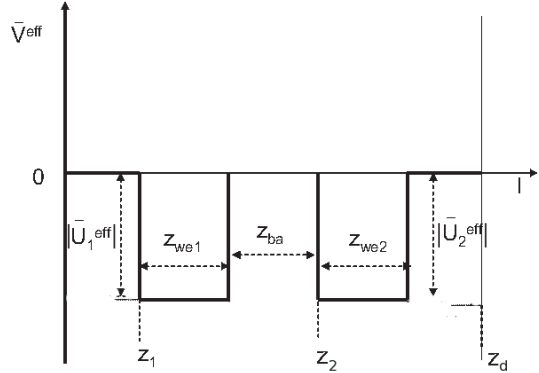,width=7cm} \caption{Model
potential for membranes without repellers. The
two wells are separated by a potential barrier of width $z_{ba}$.
$z_{1}$ ($z_{d}$) is the distance between well one (well two) and
the hard wall at $z=0$ ($z=z_d$).
The effective potential $V^{eff}=\infty$ for $z\le 0$ and $z\ge
z_{d}$.}
\end{center}
\end{figure}

\subsection{Membranes without repellers}

Let us first consider a membrane without repellers, its effective
potential is shown in Fig. 3. Since the critical phenomena for this
system belongs to Ising universality class, for sufficiently strong
potential depths the system is in two-phase state with two possible
separations $ z_{min1}$ and $ z_{min2}$ \cite{h19}; for weak
potential wells, the membrane can {\em tunnel} through the barrier
between the wells and takes one average separation field $z_{min}$.

As an example, Figure 4 shows the relation between  ${\bar
U}_{2}^{eff}$ versus $z_{min}$ given by Eq.~(\ref{eq:zMFT}) for
$z_{1}=0.1$, $z_{d}=1.2$, $z_{we1}=z_{we2}=0.2$ and $z_{ba}=0.4$.
The effective binding energy of type-1 stickers $|{\bar
U}_{1}^{eff}|$ is chosen to be $|{\bar U}_{1}^{eff}|=4$ for the
upper curve and $|{\bar U}_{1}^{eff}|= |{\bar U}_{1c}^{eff}|=1.095$
for the lower curve. One finds that $z_{min} \rightarrow z_1+
z_{we1}/2=0.2$ when $|{\bar U}_{2}^{eff}| \ll |{\bar U}_1^{eff}|$;
$z_{min}$ increases as $|{\bar U}_2^{eff}|$ increases, and $z_{min}
\rightarrow z_2+ z_{we1}/2=0.8$ when $|{\bar U}_{2}^{eff}| \gg
|{\bar U}_1^{eff}|$. However, when $|{\bar U}_1^{eff}| >|{\bar
U}_{1c}^{eff}|$ there is a range of ${\bar U}_2^{eff}$ where
$d|{\bar U}_2^{eff}|/dz_{min}$ becomes negative, this is unphysical.
The physical equation of state in the two-phase state can be found
from Maxwell's equal-area construction, which also determines the
phase boundary of the coexistence region for $|{\bar U}_1^{eff}| >
|{\bar U}_{1c}^{eff}|$. The critical point (${\bar U}_{1C}^{eff}$,
${\bar U}_{2C}^{eff}$) is found  by varying ${\bar U}_1^{eff}$ until
$\partial U_{2}^{eff} /
\partial z_{min}$ and $\partial^2 U_{2}^{eff} / \partial^2 z_{min}$
have common zero  for given $z_1$, $z_d$, $z_{ba}$, $z_{we1}$ and
$z_{we2}$.

Having discussed how to construct the phase diagram, we consider how
confinement affects this adhesion-induced lateral phase separation.
Figure 5a shows how ${\bar U}_{1c}^{eff}$ and ${\bar U}_{2c}^{eff}$
change as $z_{d}$ changes for $z_{1}=0.1$, $z_{ba}=0.4$,
$z_{we1}=0.2$ and $z_{we2}=0.2$. The effect of confinement becomes
important  for $z_{d}\le 1.5$, where $|{\bar U}_{1c}^{eff}|$ and
$|{\bar U}_{2c}^{eff}|$ increase as $z_{d}$ decreases.  This is
because  the entropic repulsion between the membrane and the hard
wall located at $z_{d}$ increases as $z_{d}$ decreases, and  the
membrane  is forced to tunnel through the barrier more often  when
$z_{d}$ decreases, thus strong confinement tend to suppress lateral
phase separation. The phase diagrams for this system at several
magnitudes of $z_d$ are shown in Fig.5b. It is clear that the
critical points shift toward greater $|U_1^{eff}|$ and $|U_2^{eff}|$
as $z_d$ decreases.  For $z_{d}=1$, the effective potential profile
shown in Fig. 3 is symmetric and the phase coexistence line is on
${\bar U}_1^{eff} = {\bar U}_{2}^{eff}$. When $z_{d}<1$, the phase
coexistence line bends up; for $z_{d}>1$ the phase coexistence line
shifts down in the vicinity of the critical point.

\begin{figure}[h]
\begin{center}
\epsfig{file=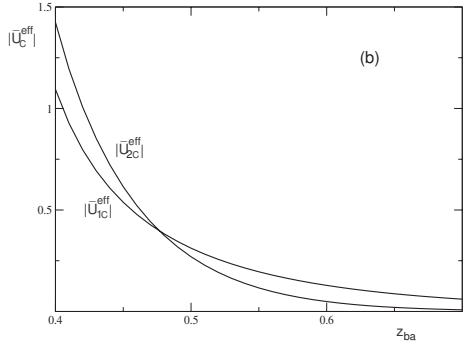,width=7cm}
 \caption{The effective
potential depth  $|{\bar U}_{2}^{eff}|$ versus $z_{min}$ for
$z_{1}=0.1$, $z_{d}=1.2$, $z_{we1}=z_{we2}=0.2$ and $z_{ba}=0.4$.
$|{\bar U}_{1}^{eff}|=4$ for the upper curve and $|{\bar
U}_{1}^{eff}|= |{\bar U}_{1c}^{eff}|=1.095$ for the lower curve,
respectively. The phase coexistence region for $|{\bar
U}_1^{eff}|=4$ can be determined  by Maxwell equal-area
construction. }
%\label{figfactor}
\end{center}
\end{figure}
\begin{figure}
\centering
%\subfigure[Bild a.] % caption for subfigure a
{
    \label{fig:sub:a}
    \includegraphics[width=6cm]{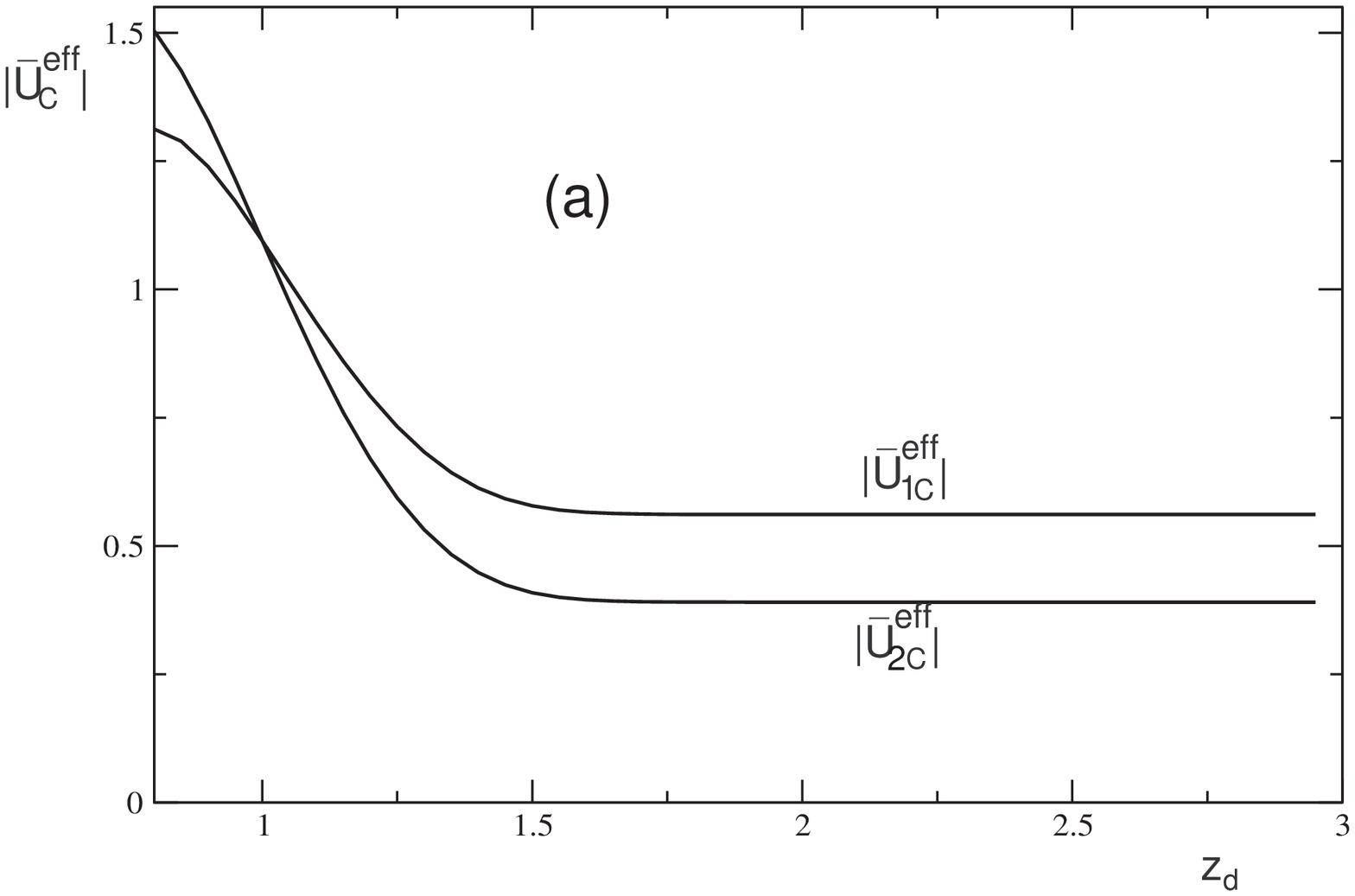}
}
\hspace{1cm}
%\subfigure[Bild b.] % caption for subfigure b
{
    \label{fig:sub:b}
    \includegraphics[width=6cm]{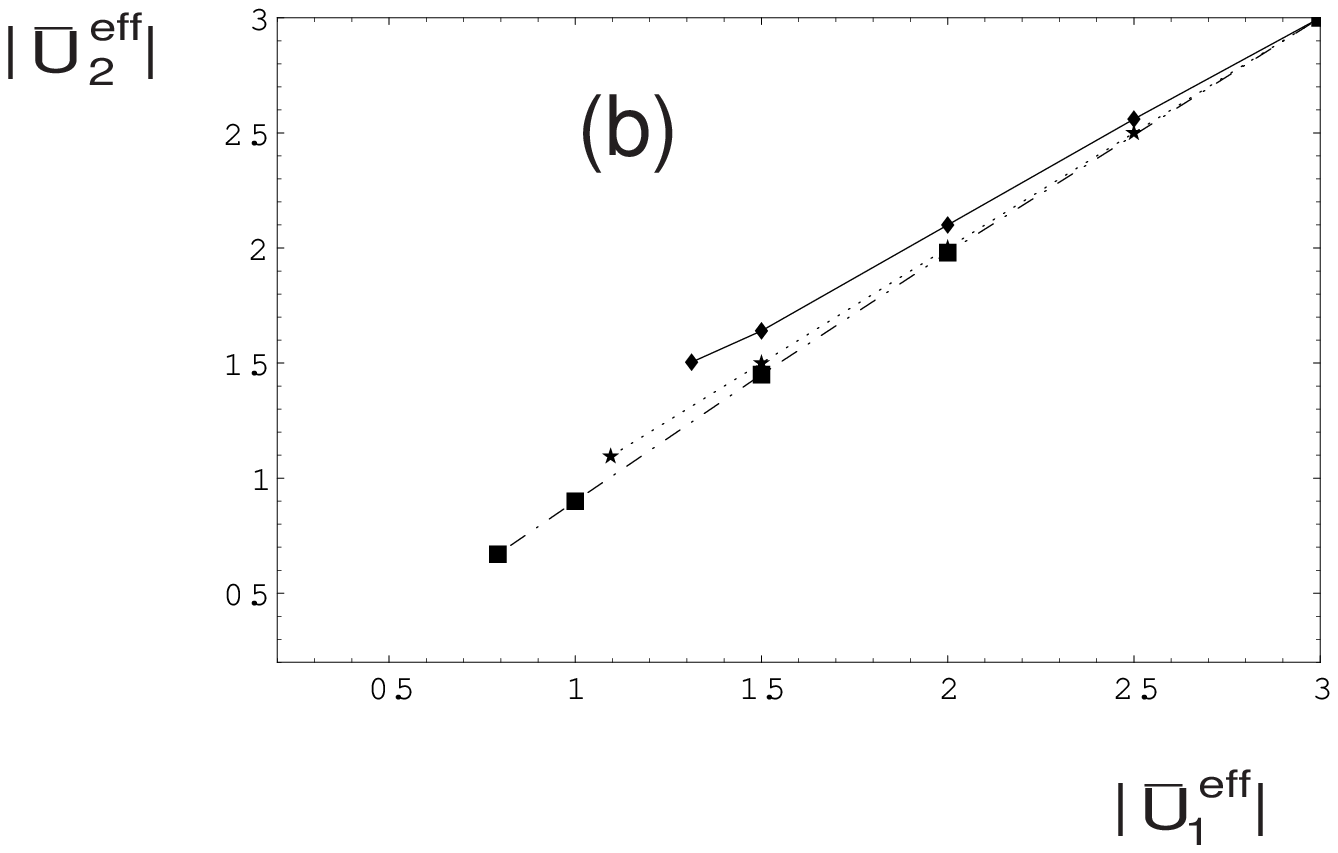}
} \caption{(a) The critical potential depths versus $z_{d}$ for
$z_{1}=0.1$, $z_{ba}=0.4$, $z_{we1}=0.2$ and $z_{we2}=0.2$. When
$z_{d}$ is small, $|{\bar U}_{1c}^{eff}|$ and $|{\bar
U}_{2c}^{eff}|$ increases as $z_{d}$ decreases. (b) The phase
boundaries for the system shown in Fig.5a.  $z_{d}=0.9$ (top),
$z_{d}=1$ (middle), and $z_{d}=1.2$ (bottom). }
\label{fig:sub} % caption for the whole figure
\end{figure}

\begin{figure}
\centering
%\subfigure[Bild a.] % caption for subfigure a
{
    \label{fig:sub:a}
    \includegraphics[width=6cm]{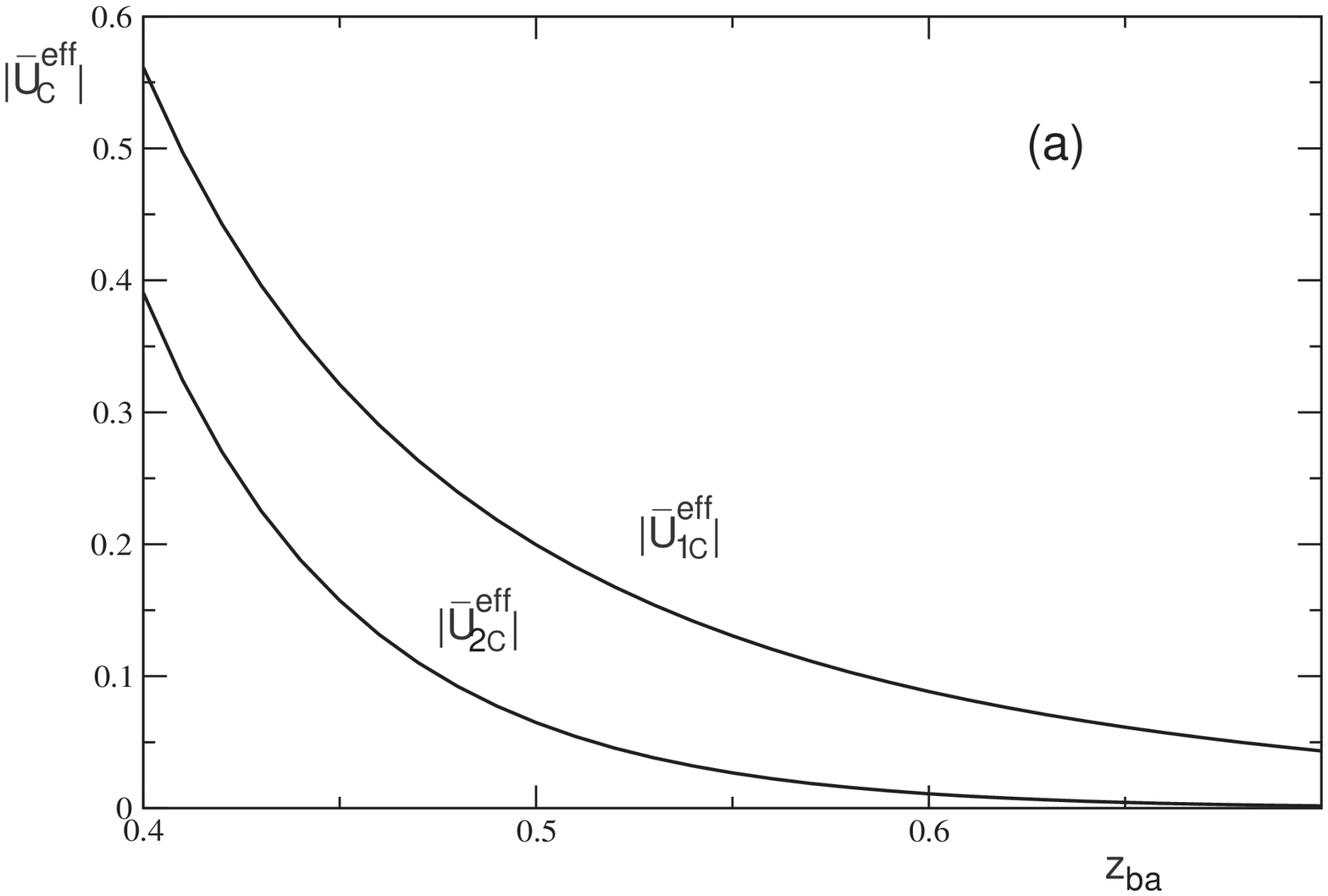}
}
\hspace{1cm}
%\subfigure[Bild b.] % caption for subfigure b
{
    \label{fig:sub:b}
    \includegraphics[width=6cm]{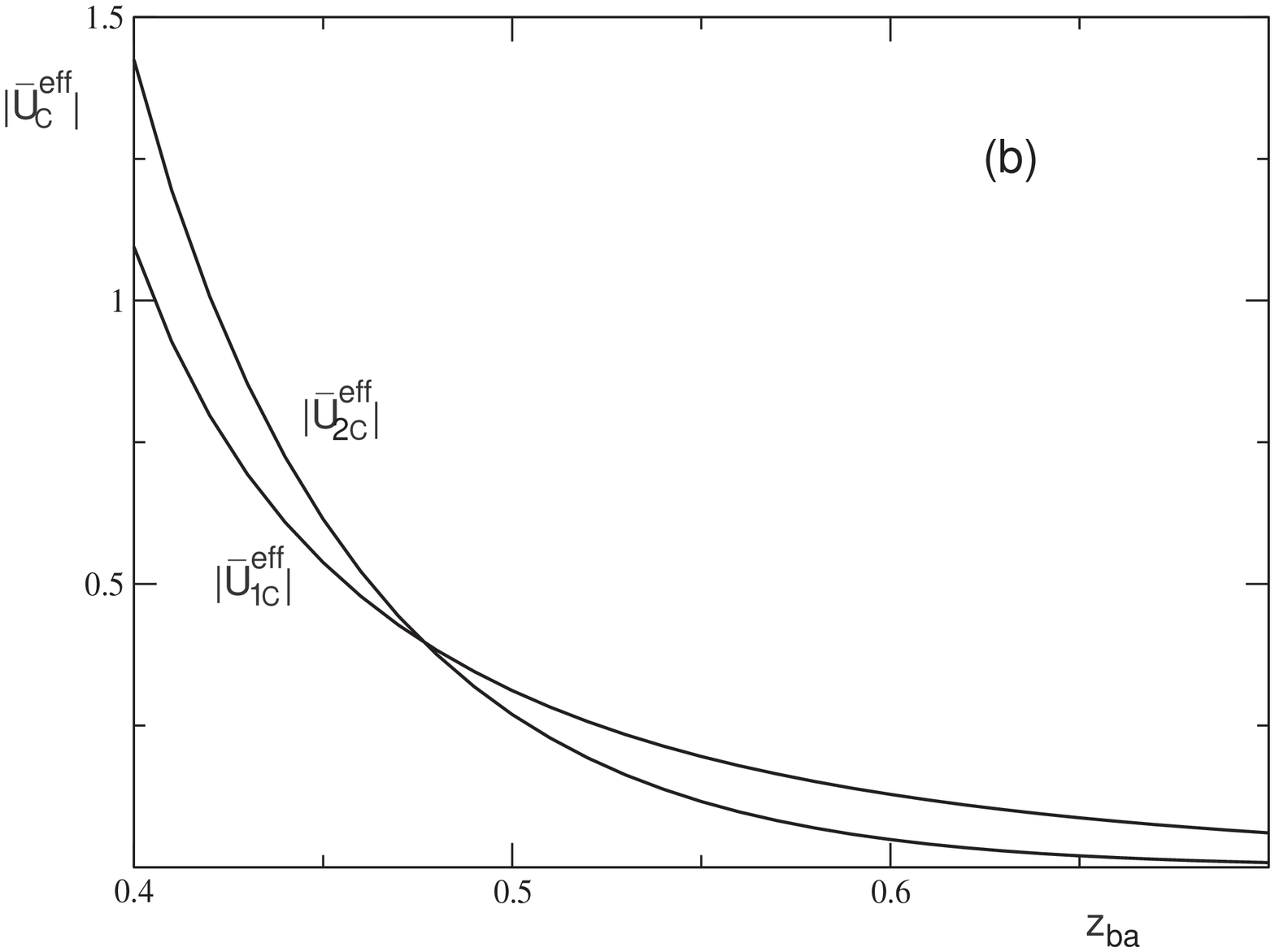}
} \caption{(a) The critical potential depths versus $z_{ba}$ for
$z_{1}=0.1$, $z_{d}=6$, and $z_{we1}=z_{we2}=0.2$. $|{\bar
U}_{1c}^{eff}|>|{\bar U}_{2c}^{eff}|$ for all $z_{ba}$ and both
$|{\bar U}_{1c}^{eff}|$, $|{\bar U}_{2c}^{eff}|$ decrease as
$z_{ba}$ increases. (b) The critical potential depths versus
$z_{ba}$ for $z_{1}=0.1$, $z_{d}=6$, $z_{we1}=0.2$, and
$z_{we2}=0.1$. $|{\bar U}_{2c}^{eff}|>|{\bar U}_{1c}^{eff}|$  for
small $z_{ba}$, and $|{\bar U}_{1c}^{eff}|>|{\bar U}_{2c}^{eff}|$
for large $z_{ba}$.}
\label{fig:sub} % caption for the whole figure
\end{figure}
Mean field theory is also convenient for studying how ${\bar
U}_{1c}^{eff}$ and ${\bar U}_{2c}^{eff}$ vary as the length
difference between the stickers changes. Fig.6a shows that, when the
effect of confinement is negligible, as the length difference
between short and long stickers increases, lateral phase separation
occurs at lower $|{\bar U}_{1c}^{eff}|$ and $|{\bar U}_{2c}^{eff}|$,
as one expected. Furthermore, for $z_{we1}=z_{we2}$, $|{\bar
U}_{1c}^{eff}| > |{\bar U}_{2c}^{eff}|$ due to collisions between
the membrane and the substrate.  On the other hand, Fig.6b depicts
that when $z_{we2}= {1/2}z_{we1}$, $|{\bar U}_{2c}^{eff}|>|{\bar
U}_{1c}^{eff}|$ when $z_{ba}$ is small due to the potential width
difference.  However, for large $z_{ba}$ the steric repulsion
between a membrane in the second well and the substrate becomes
unimportant, thus $|{\bar U}_{2c}^{eff}|<|{\bar U}_{1c}^{eff}|$ even
though $z_{we2}< z_{we1}$.  These results demonstrate that our mean
field theory can be applied to analyze various physical effects on
the adhesion-induced lateral phase separation.  A more detailed
study will likely require time-consuming large-scale numerical
simulations.

To check whether the physics revealed by our simple mean-field
analysis holds when fluctuations are taken into account, we compare
the mean-field result with Monte Carlo simulation.

When $z_{d}$ and $z_{1}$ are both large, the effect of the walls is
negligible. Hence when $z_{we1}=z_{we2}$, the membrane is
effectively in a symmetric double-well potential \cite{h18}.
  The critical potential depths ${\bar U}_{1c}^{eff} = {\bar U}_{2c}^{eff }={\bar
U}^{eff}$. The location of the critical potential depth ${\bar
U}_{c}^{eff }$ can be obtained  by using Binder cumulant method
\cite{h28}. The   ${\bar U}^{eff}$ dependence of  the Binder moments
$C_{2}=\left\langle {\bar z}^2 \right\rangle /\left\langle |{\bar
z}| \right\rangle ^2 $ and  $ C_{4}=\left\langle {\bar z}^4
\right\rangle /\left\langle {\bar z}^2 \right\rangle ^2 $  is
calculated for several system sizes and the critical point is
located at the common intersection point of  those curves due to the
divergence of the correlation length at criticality. ${\bar
z}={1\over N }\sum_{i=1}^{N}$ denotes the spatial average of the
separation field while $\left\langle ... \right\rangle $ represents
thermal average. As an example, Figures 7a and 7b show  $C_{2}$ and
$C_{4}$ versus ${\bar U}^{eff }$ for  $z_{1}=0.8$, $z_{d}= 3.1$,
$z_2=1.8$, $z_{we1}=z_{we2} =0.5$, and $z_{ba}=0.5$. The system size
$L \times L$ in the simulations are $L=10$, $L=20$ and $L=30$.
  ${\bar U}_{c}^{eff}$ can be  obtained from the  common
intersection point of $C_{2}$ and $C_{4}$  for different $L$.

The effect of confinement on lateral phase separation of membrane is
important when $z_{d}$ and $z_{1}$ are small.  In this case the
walls affect the phase coexistence line and the critical potential
depths. Thus the phase coexistence line and the critical potential
depths in the simulations are determined by measuring the binding
probability $P_1$ of the membrane in well-one, and the binding
probability $P_2$ of the membrane in well-two. The simulation starts
in the regime when both potential wells are deep and the membrane
stays in well-two; then we decrease $|{\bar U}_{2}^{eff}|$, $P_2$
decreases continuously until the membrane switches from well-two to
well-one. This discontinuous transition signals a first-order phase
transition\cite{mes}. The location of the critical point can be
determined by repeating the above procedure for systems with
progressively smaller $|{\bar U}_{1}^{eff}|$. Below the critical
point, the plot $P_{2}$ versus ${\bar U}_{2}^{eff}$ is continuous.

Figure 8 shows the phase diagram for $z_{1}=0.1$, $z_{ba}=0.4$,
$z_{we1}=0.2$, and $z_{we2}=0.2$.
%The parameter $z_{d}$ is chosen to
%be $z_{d}=0.9$, $z_{d}=1$ and $z_{d}=1.2$ form up to down.
For small $z_{d}$, the phase coexistence curve shifts up in the
vicinity of the critical potential depth as membrane confined in
well two feels higher entropic repulsion with the hard wall than
membrane confined in well-one. For $z_{d}=1$, the effective
potential is symmetric thus the phase coexistence line is at ${\bar
U}_{1}^{eff}={\bar U}_{2}^{eff}$. %and the critical potential depth is
%obtained using cummulant method.
For $z_d =1.2$ the phase boundary shifts down in the vicinity of the
critical point.  Although the critical points in the simulations are
located at higher $|U_{1}^{eff}|$ and $|U_2^{eff}|$ than those in
the mean field theory due to fluctuations, simulations also shown
confinement enhanced phase separation. Thus, although mean field
theory cannot provide accurate prediction for the critical points,
it provides good prediction about the shape of phase boundary in
$|U_1^{eff}|-|U_2^{eff}|$ plane and the entropic effect of
confinement on the phase boundary.
\begin{figure}
\centering
%\subfigure[Bild a.] % caption for subfigure a
{
    \label{fig:sub:a}
    \includegraphics[width=6cm]{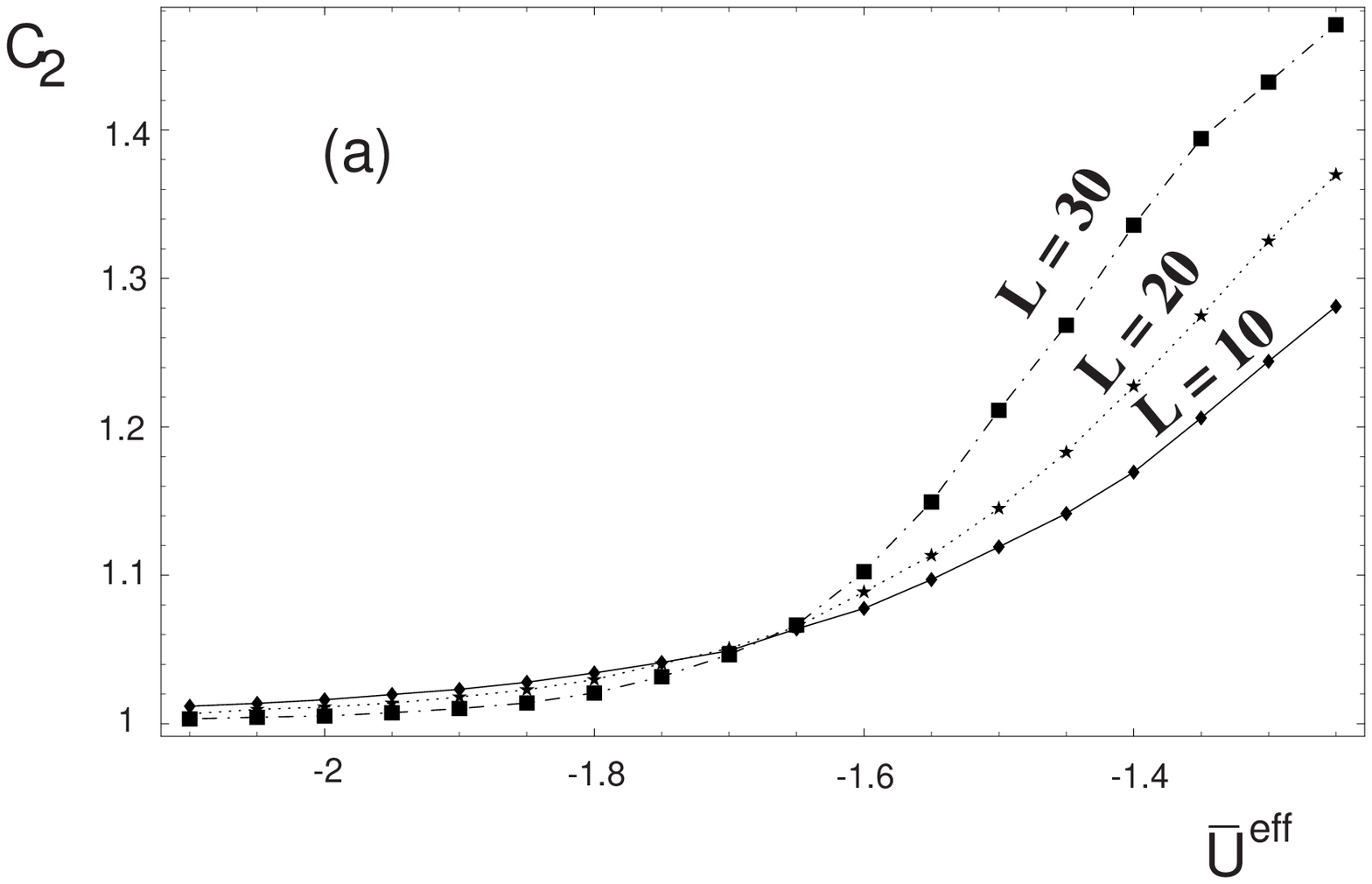}
}
\hspace{1cm}
%\subfigure[Bild b.] % caption for subfigure b
{
    \label{fig:sub:b}
    \includegraphics[width=6cm]{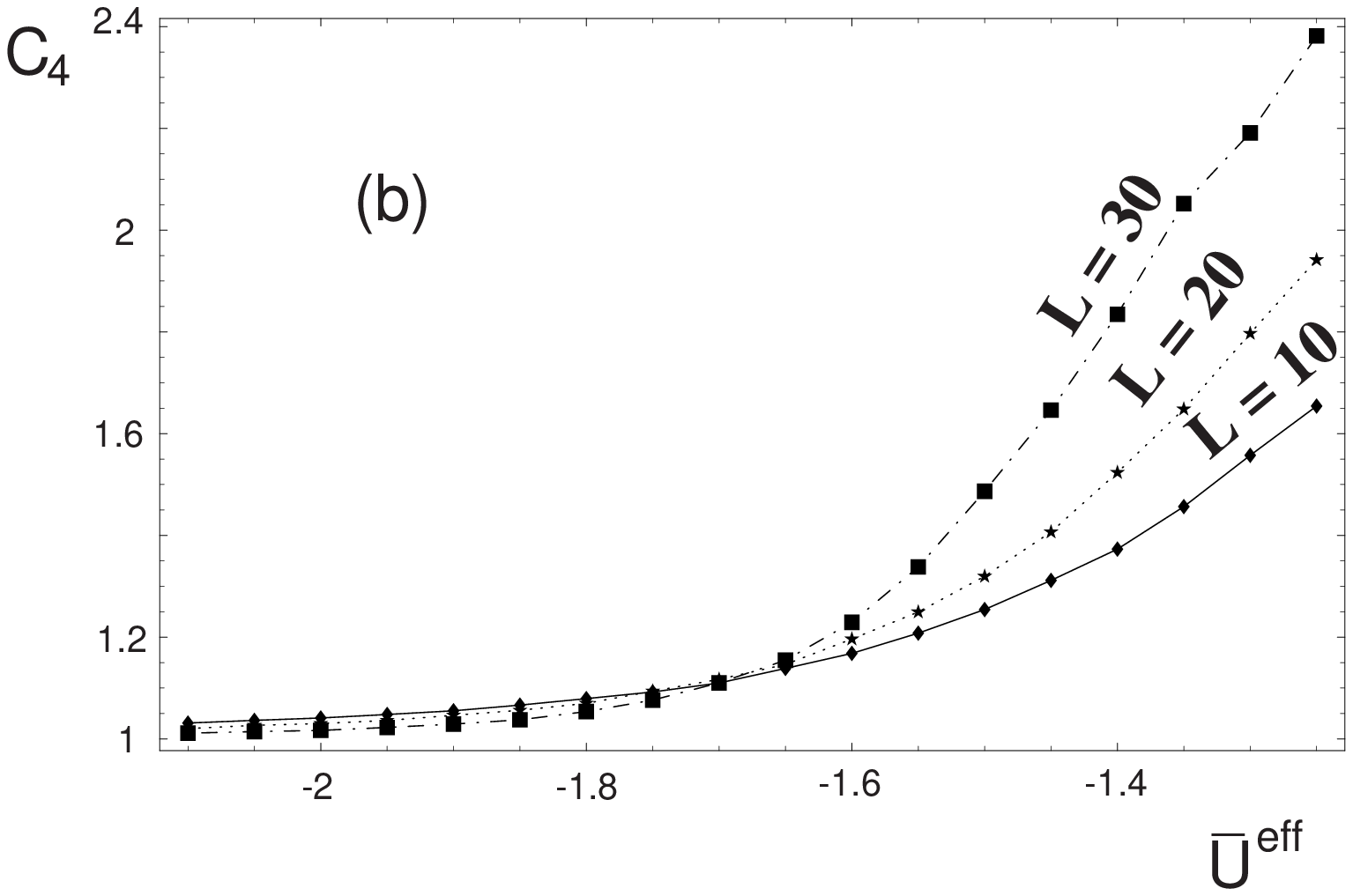}
} \caption{(a) The cummulant $C_{2}$ versus ${\bar U}^{eff}$ for
$z_{1}=0.8$, $z_{d}= 3.1$, $z_2=1.8$, $z_{we1}=z_{we2} =0.5$, and
$z_{ba}=0.5$. The intersection point for $L=10$, $L=20$, and $L=30$
denotes the location of the critical potential depth ${\bar
U}_{C}^{eff}$. (b) The cummulant $C_{4}$ versus ${\bar U}^{eff}$ for
$z_{1}=0.8$, $z_{d}= 3.1$, $z_2=1.8$, $z_{we1}=z_{we2} =0.5$ and
$z_{ba}=0.5$. }
\label{fig:sub} % caption for the whole figure
\end{figure}

\begin{figure}[h]
\begin{center}
\epsfig{file=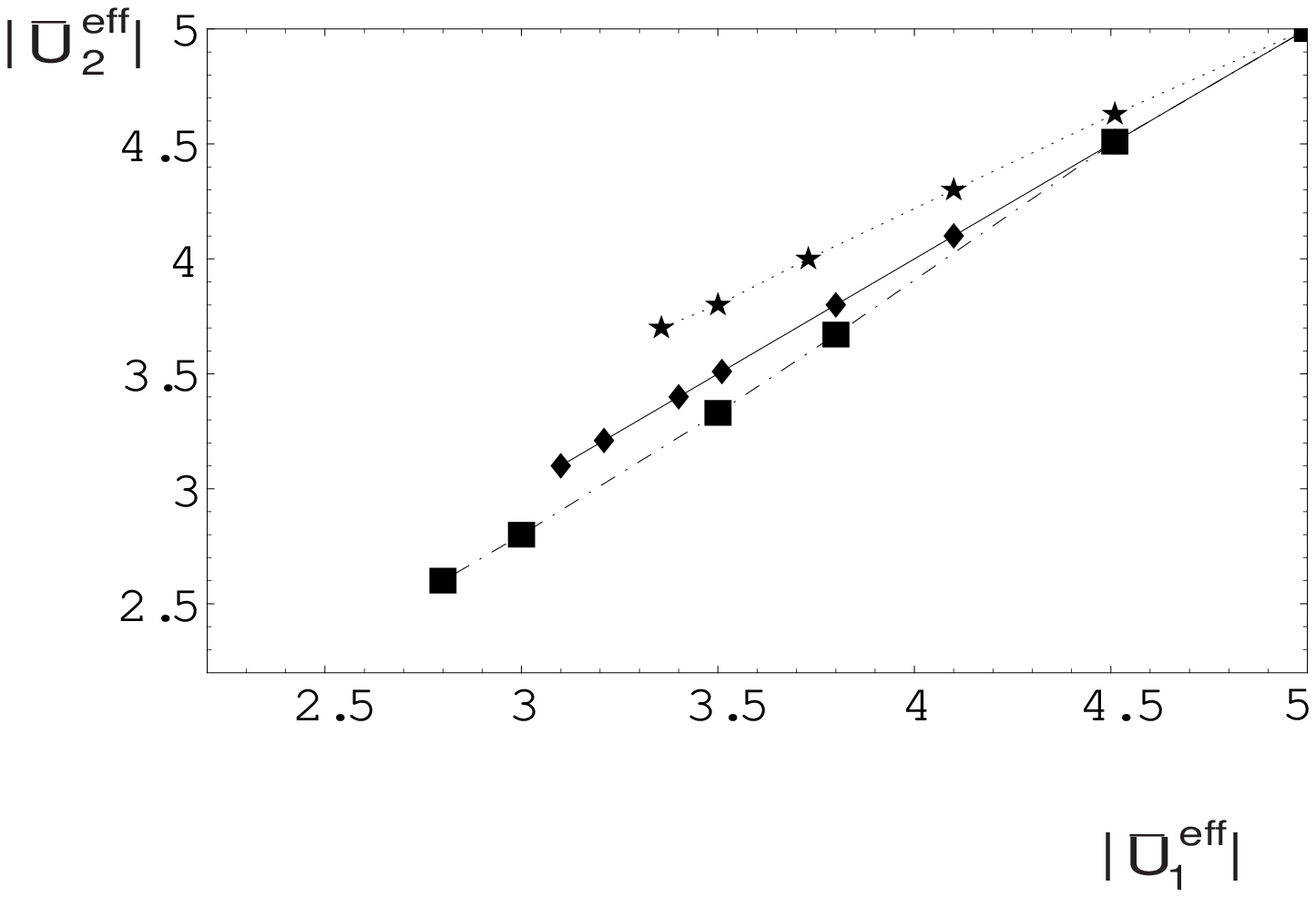,width=7cm}
 \caption{ Phase diagram in
${\bar U}_{1}^{eff}$ ${\bar U}_{2}^{eff}$ space constructed from
Monte Carlo simulations for membranes without repellers.  In the
simulations $z_{1}=0.1$, $z_{ba}=0.4$, $z_{we1}=0.2$, and
$z_{we2}=0.2$.  $z_{d}=0.9$ (top), $z_{d}=1$ (middle), and
$z_{d}=1.2$ (bottom). }
%\label{figfactor}
\end{center}
\end{figure}

\subsection{Membranes with  repellers}

   To study the effect of repellers on adhesion-induced lateral
phase separation, first notice that adding repellers to a system
means for given sticker species and densities ($U_1$, $\mu _1$,
$U_2$, and $\mu _2$ are not changed), repellers with given $U_3$,
$\mu _3$, and $l_3$ are added to the system.
%To study whether the
%presence of repellers tend to enhance or suppress lateral phase
%separation of the stickers during membrane adhesion,
Therefore we need to see how effective potentials associated with
the stickers change as repellers are added to the system.

   For membranes containing repellers, the
effective potential of the membrane takes  the form
\begin{eqnarray}
{\bar V}^{eff} = \left\{ \begin{array}{ll}
  \infty , &\mbox{for}~ z<0, \\
  {\bar U}_{ba}^{eff}, &\mbox{for}~ 0<z<z_{1},  \\
  {\bar U}_{1}^{eff}\equiv [{\bar U}_1^{eff}]_{trans}, &\mbox{for} ~z_{1}<z<z_{1}+z_{we1},  \\
  {\bar U}^{eff}_{ba}, &\mbox{for}~ z_{1}+z_{we1}<z<z_{2},  \\
  {\bar U}_{2}^{eff} \equiv [{\bar U}_2^{eff}]_{trans}, &\mbox{for}~ z_{2}<z<z_{2}+z_{we2}, \\
  {\bar U}_{ba}^{eff}, &\mbox{for}~ z_{2}+z_{we2}<z<z_{3}, \\
  0, &\mbox{for} ~z_{3}<z<z_{d},\\
  \infty, &\mbox{for}~ z>z_{d}
  \end{array} \right.
\label{eq:U_effrep}
\end{eqnarray}
as shown in Fig.9. Here ${\bar U}_{1}^{eff}<0$ and ${\bar
U}_{2}^{eff}<0$ while ${\bar U}_{ba}^{eff}>0$.
\begin{figure}[h]
\begin{center}
\epsfig{file=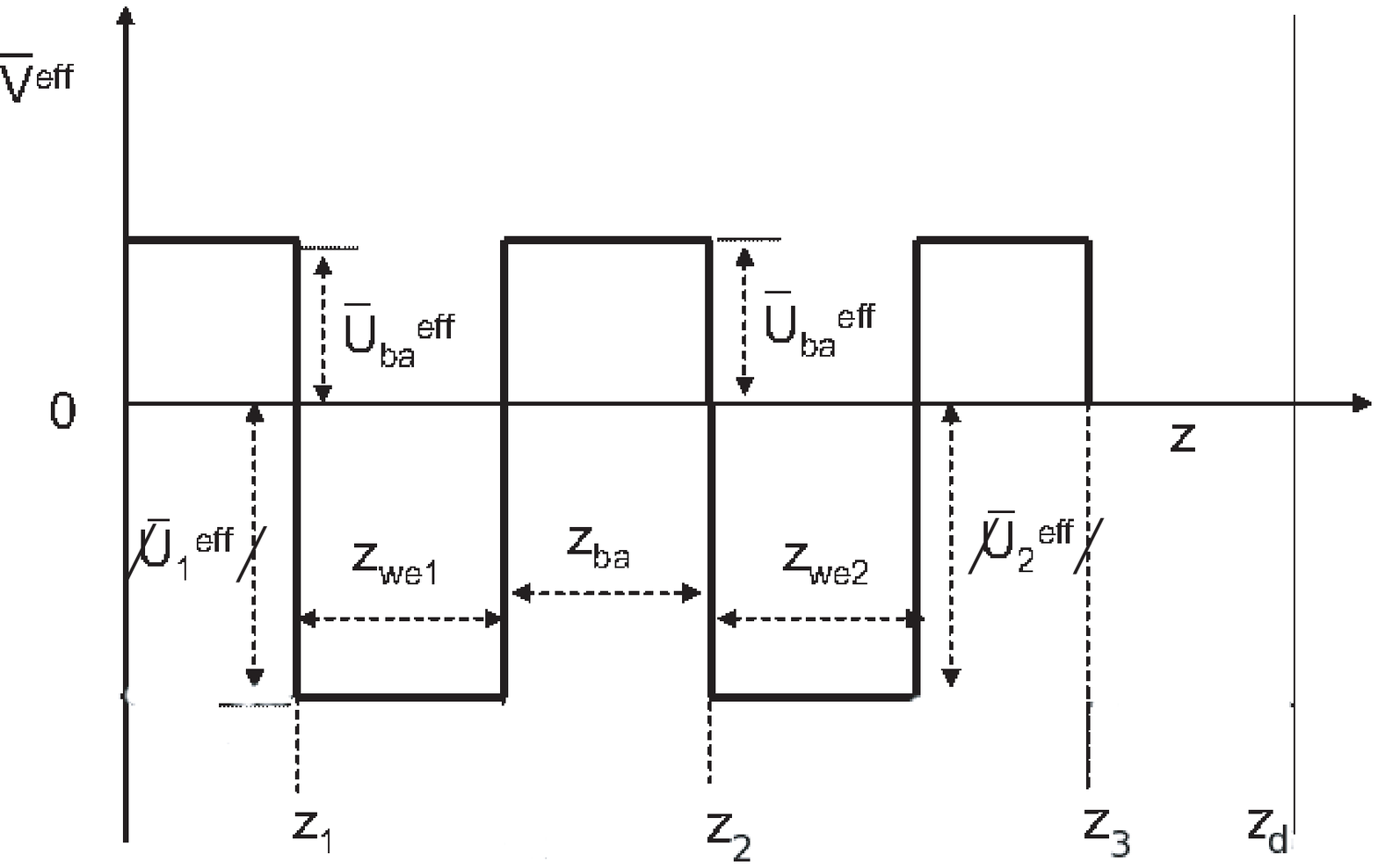,width=7cm} \caption{A model potential with two
square wells of depth $|{\bar U}_{1}^{eff}|$ and $|{\bar
U}_{2}^{eff}|$ within the range $z_{we1}$ and  $z_{we2}$, and one
barrier of height ${\bar U}_{ba}^{eff}$ and width $z_{ba}$. Because
of the hard walls, the effective potential $V^{eff}=\infty$ for
$z\le 0$ and $z\ge z_{d}$.}
%\label{figfactor}
\end{center}
\end{figure}
The presence of repellers contribute the effective potentials of the
stickers in Eq.~(\ref{eq:U_effrep}).  To make this point more
transparent, let the effective potential of sticker-$i$ ($i=1$ or 2)
in the absence of repellers be
\begin{eqnarray}
 [{\bar U}_1^{eff}]_0
= - T \ln \frac{1+e^{(-U_1+\mu _1)/T}+ e^{\mu _2/T}}{1+e^{\mu
_1/T}+e^{\mu _2/T}}, \label{eq:U_10}
\end{eqnarray}
and
\begin{eqnarray}
[{\bar U}_2^{eff}]_0 = - T \ln \frac{1+e^{\mu _1/T}+ e^{(-U_2+\mu
_2)/T}}{1+e^{\mu _1/T}+e^{\mu _2/T}}. \label{eq:U_20}
\end{eqnarray}
In the presence of repellers, the effective potentials become
\begin{eqnarray}
  [{\bar U}_1^{eff}]_{trans}
= - T \ln \frac{1+e^{(-U_1+\mu _1)/T}+ e^{\mu _2/T}+e^{\mu
_3/T}}{1+e^{\mu _1/T}+e^{\mu _2/T}+e^{\mu _3/T}}, \label{eq:U_1rep}
\end{eqnarray}
\begin{eqnarray}
  [{\bar U}_2^{eff}]_{trans}
= - T \ln \frac{1+e^{\mu _1/T}+ e^{(-U_2+\mu _2)/T}+e^{\mu
_3/T}}{1+e^{\mu _1/T}+e^{\mu _2/T}+e^{\mu _3/T}}, \label{eq:U_2rep}
\end{eqnarray}
and the effective potential of the repellers is
\begin{eqnarray}
{\bar U}_{ba}^{eff} = - T \ln \frac{1+e^{\mu _1/T}+ e^{\mu
_2/T}+e^{(-U_3 + \mu _3)/T}}{1+e^{\mu _1/T}+e^{\mu _2/T}+e^{\mu
_3/T}} \label{eq:Uba}
\end{eqnarray}
 Intuitively, adding
repellers to the system reduces the affinity of the stickers, this
can verified by straightforward algebra. Indeed, from
Eqs.~(\ref{eq:U_10})(\ref{eq:U_20})(\ref{eq:U_1rep})(\ref{eq:U_2rep}),
one finds
\begin{eqnarray}
|[{\bar U}_1^{eff}]_{trans}|&=&|[{\bar U}_1^{eff}]_0|+T \ln
\frac{1+e^{\mu _3/T}/\left[1+e^{(-U_1+\mu _1)/T}+e^{\mu
_2/T}\right]}{1+e^{\mu _3/T}/(1+e^{\mu
_1/T}+e^{\mu _2/T})} \nonumber \\
&<& |[{\bar U}_1^{eff}]_0|, \label{eq:U_1trans}
\end{eqnarray}
and
\begin{eqnarray}
|[{\bar U}_2^{eff}]_{trans}|&=&|[{\bar U}_2^{eff}]_0|+T \ln
\frac{1+e^{\mu _3/T}/\left[1+e^{\mu _1/T}+e^{(-U_2+\mu
_2)/T}\right]}{1+e^{\mu _3/T}/(1+e^{\mu
_1/T}+e^{\mu _2/T})} \nonumber \\
&<& |[{\bar U}_1^{eff}]_0|, \label{eq:U_2trans}
\end{eqnarray} because $U_1<0$, and $U_2<0$.

The above discussion suggests that to see if the presence of
repellers enhances or suppresses adhesion-induced lateral phase
separation of different species of stickers, one needs to compare
the critical potentials $|{\bar U}_{ic}^{eff}|$ in the presence of
repellers with $|[{\bar U}_{ic}^{eff}]_{trans}|$, the potentials
that are transformed from $|[{\bar U}_{ic}^{eff}]_0|$ by
Eqs.~(\ref{eq:U_1trans})(\ref{eq:U_2trans}).

As demonstrated in the previous section, although quantitatively not
accurate, mean field approximation gives us correct physical picture
of the system under consideration. Since the precise magnitude of
the critical potential depths is not the key issue of this section,
we use mean field theory to study the effect of repellers. First we
check if the effect of confinement in the presence of repellers is
the same as that in the absence of repellers.  The critical
potential depths in the presence of repellers versus $z_{d}$ for
$z_{1}=0.1$, $z_{3}=1.0$, $z_{ba}=0.4$, $z_{we1}=0.2$,
$z_{we2}=0.2$, and $|{\bar U}_{ba}^{eff}|=0.1$ in the mean field
approximation are shown in Fig.10a. Indeed, like the no-repeller
case, the the critical potential depths $|{\bar U}_{1c}^{eff}|$ and
$|{\bar U}_{2c}^{eff}|$ decrease as $z_{d}$ increases. Furthermore,
the confinement effect is negligible for large values of $z_{d}$,
this is also the same as no-repeller case.
\begin{figure}
\centering
%\subfigure[Bild a.] % caption for subfigure a
{
    \label{fig:sub:a}
    \includegraphics[width=6cm]{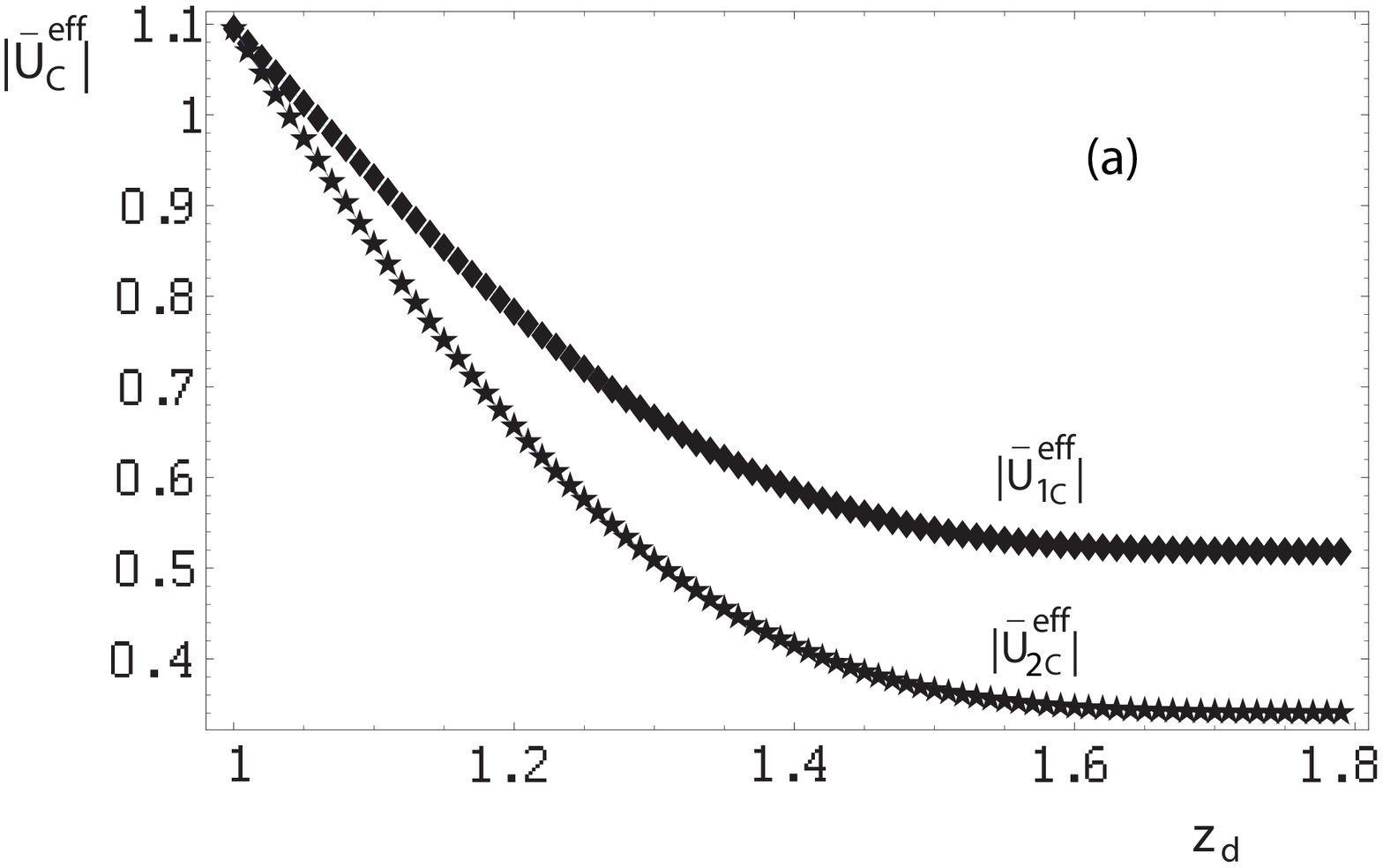}
}
\hspace{1cm}
%\subfigure[Bild b.] % caption for subfigure b
{
    \label{fig:sub:b}
    \includegraphics[width=6cm]{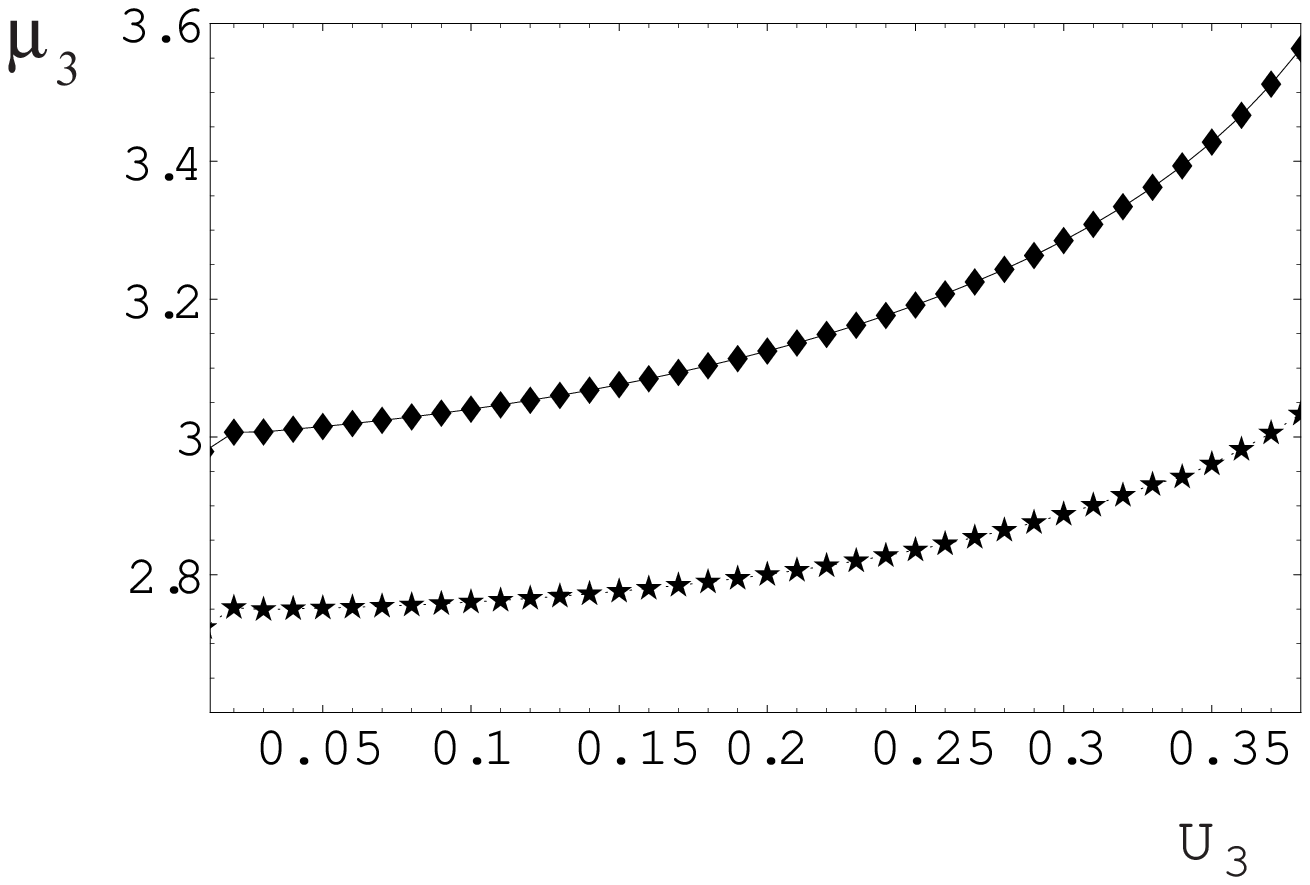}
} \caption{(a) The critical potential depths versus $z_{d}$ for
$z_{1}=0.1$, $z_{3}=1.0$, $z_{ba}=0.4$, $z_{we1}=0.2$,
$z_{we2}=0.2$, and $|{\bar U}_{ba}^{eff}|=0.1$.  For small $z_{d}$,
$|{\bar U}_{c}^{eff}|$  decreases  as $z_{d}$ increases. %
(b) Upper curve: $|{\bar U}_{2c}^{eff}|-|[{\bar
U}_{2c}^{eff}]_{trans}|=0$, lower curve: $|{\bar
U}_{1c}^{eff}|-|[{\bar U}_{1c}^{eff}]_{trans}|=0$.  On the left of
the curves, repellers suppress lateral phase separation; on the
right of the curves, repellers enhance lateral phase separation;
between the curves, $|{\bar U}_{2c}^{eff}|-|[{\bar
U}_{2c}^{eff}]_{trans}|<0$, and $|{\bar U}_{1c}^{eff}|-|[{\bar
U}_{1c}^{eff}]_{trans}|>0$.  The curves are plotted for $z_{1}=0.1$,
$z_{3}=1$, $z_{d}=2$, $z_{we1}=0.2$, $z_{we2}=0.2$, and
$z_{ba}=0.4$.}
\label{fig:sub} % caption for the whole figure
\end{figure}

    To see the effect of repellers on adhesion-induced phase separation,
we compare the critical potentials of the stickers $|[{\bar
U}_{ic}^{eff}]|$ with $|[{\bar U}_{ic}^{eff}]_{trans}|$. Figure 10b
shows the curves on which $|{\bar U}_{ic}^{eff}|=|[{\bar
U}_{ic}^{eff}]_{trans}|$ for $z_{1}=0.1$, $z_{d}=6$,
$z_{we1}=z_{we2}=0.2$, and $z_{ba}=0.4$. Repellers suppress phase
separation on the left of the curves, and enhance phase separation
on the right of the curves.  Between the curves $|{\bar
U}_{2c}^{eff}|-|[{\bar U}_{2c}^{eff}]_{trans}|<0$, and $|{\bar
U}_{1c}^{eff}|-|[{\bar U}_{1c}^{eff}]_{trans}|>0$.  This indicates
that stiff repellers enhance, while soft repellers suppress
adhesion-induced lateral phase separation.  This result can be
understood by a simple analysis. When membrane-membrane collisions
is not important, adding repellers should not significantly change
the height of energy barrier between the wells of the stickers, thus
the critical potentials of the stickers in the presence of repellers
are related to those in the absence of repellers by $|{\bar
U}_{ic}^{eff}|+{\bar U}_{ba}^{eff}\approx |[{\bar
U}_{ic}^{eff}]_0|$. Repellers enhance phase separation as long as
$|[{\bar U}_{ic}^{eff}]_{trans}| > |{\bar U}_{ic}^{eff}|\approx
|[{\bar U}_{ic}^{eff}]_0|-{\bar U}_{ba}^{eff}$. From
Eqs.~(\ref{eq:U_1trans})(\ref{eq:U_2trans})(\ref{eq:Uba}), this
condition leads to
\begin{eqnarray}
T \ln \left[ \frac{1+e^{\mu _3/T}/(1+e^{(-U_1+\mu _1)/T}+e^{\mu
_2/T})}{1+e^{(-U_3+\mu _3)/T}/(1+e^{\mu _1/T}+e^{\mu _2/T})} \right]
> 0
\end{eqnarray}
for type-1 stickers.  Since $U_3>0$, we find that for sufficiently
large $U_3$ (i.e., stiff repellers) the above inequality is
satisfied and phase separation is enhanced.  Similarly, when $U_3$
is sufficiently large, $|{\bar U}_{2c}^{eff}|-|[{\bar
U}_{2c}^{eff}]_{trans}|<0$.

\section{Summary and conclusion}

%The effect of vertical confinement on lateral phase separation of
%membranes with short and long stickers is explored  at  mean field
%level and with Monte Carlo simulations.   The result obtained in
%this work shows that for large values of $z_{d}$ the effect of
%confinement is negligible while for smaller values of $z_{d}$, the
%critical potential depths   $|{\bar U}_{1c}^{eff}|$  and $|{\bar
%U}_{2c}^{eff}|$  increase as $z_{d}$ decreases.
%Due to the asymmetry in the potential profile, the critical
%potential depths ${\bar U}_{1c}^{eff}$ and ${\bar U}_{2c}^{eff}$
%may not be equal. ${\barU}_{1c}^{eff}$ and ${\bar U}_{2c}^{eff}$
%decrease as the parameter $z_{ba}$ increases.
%When the width of the two potential wells is equal
%$z_{we1}=z_{we2}$, the critical potential depth $|{\bar
%U}_{1c}^{eff}| > |{\bar U}_{2c}^{eff}|$ as membranes confined in the
%first-well experience higher steric repulsion with hard wall. On the
%other hand when $z_{we2}<z_{we1}$, membranes confined in the
%second-well feel higher entropic loss due to confinement in the
%potential well and hence $|{\bar U}_{2c}^{eff}| > |{\bar
%U}_{1c}^{eff}|$ for smaller value of $z_{ba}$. However, for large
%$z_{ba}$,  $|{\bar U}_{1c}^{eff}| > |{\bar U}_{2c}^{eff}|$ as the
%steric repulsion of membrane with hard wall for membrane confined in
%well-one is higher than the membrane confined in well-two. At a
%certain $z_{ba}$, $|{\bar U}_{1c}^{eff}| = |{\bar U}_{2c}^{eff}|$.

We have developed a mean field analysis that is convenient for
studying the phase behavior of membrane adhesion induced lateral
phase separation.  Our study shows that vertical confinement tends
to suppress adhesion-induced phase separation because
long-sticker-rich state is suppressed due to the entropic loss.  We
also find that adding repellers reduce the effective binding
energies of the stickers, and repellers play a non-trivial role in
adhesion-induced phase separation: stiff repellers tend to enhance
phase separation, soft repellers tend to suppress phase separation.
These ideas are not difficult to check in experiments. For example,
consider vesicle adhesion to supported membranes via two types of
stickers. Our analysis predicts that it is possible to mix the
phase-separated stickers by simply compressing the vesicle against
the supporting substrate.  The effect of repellers can be checked by
incorporating non-adhesive flexible polymers and stiff rod-like
molecules to the vesicle surface, and examine the adhesion zone.
Flexible polymers should suppress lateral phase separation while
stiff molecules should enhance lateral phase separation.  We believe
that these effects could be useful in the development of new
sensitive soft materials with possible applications in future
bio-technologies.
%The critical potential depths $|{\bar U}_{1c}^{eff}|$ and $|{\bar U}_{2c}^{eff}|$
%decrease as $z_{we1}$, $z_{we2}$, $z_{ba}$, and  $z_{1}$ increase.
%The repeller potential height also significantly affects the phase
%property of the system. The result obtained in this theoretical work
%shows that as the height of the repellers potential increases, the
%critical potential depths $|{\bar U}_{1c}^{eff}|$ and $|{\bar
%U}_{2c}^{eff}|$ decrease.

% In this work we study adhesion-induced lateral phase separation of
%membranes with short, long stickers and repellers with mean-field
%approximation and with Monte Carlo simulations.   The effect of
%confinement on  lateral phase separation of multi-component
%membranes is explored.  The result obtained in this theoretical work
%shows that confinement facilities mixing of stickers and repellers
%and, its effect is significant at smaller values of $z_{d}$.

\section*{\bf Acknowledgement}
MA would like to thank Prof. R. Lipowsky,  T. R. Weikl and  B.
Rozycki for discussions he had  during his stay at  Max Planck
institute for Colloids and Interfaces, Potsdam, Germany. MA  would
like  also to thank   Mulugeta Bekele for the interesting
discussions.  This work is supported by National Science Council of
Taiwan, Republic of China, Grant No. NSC 96-2628-M-008 -001 -MY2.

\end{document}